# Agent-Based Model Framework for the North Carolina Modeling Infectious Diseases Program (NC MInD ABM)

# Overview, Design Concepts, and Details Protocol

Kasey Jones[1], Emily Hadley[1], Caroline Kery[1], Alexander Preiss[1], Marie C. D. Stoner[1], Sarah Rhea[2]

[1]RTI International, Research Triangle Park, North Carolina
[2]North Carolina State University, Department of Veterinary Medicine, Raleigh, North Carolina

## Abstract

To help facilitate a variety of simulations related to healthcare facilities in North Carolina, we have developed an agent-based model (ABM) to accurately simulate patient (i.e., agent) movement to and from these facilities. This is an Overview, Design Concepts, and Details (ODD) Protocol, a standardized method for describing ABMs. This ODD provides detailed information on healthcare facilities in North Carolina, the agent movement to and between them, and any decisions that were made during the creation of this model. This ABM is intended to be used alongside disease-specific submodels. It can be used for purposes such as simulating the success of interventions on reducing disease transmission, simulating strain on facility resources (including staff and materials), and forecasting hospital capacity. Disease-specific ODDs should accompany this document. No details related to any submodels that use this ABM as a base model are included.

Human subjects' statement: This work was approved by the Institutional Review Board at RTI International.

Funding Sources: North Carolina Department of Health and Human Services Contract #2020-DHHS-COVPS 010; Centers for Disease Control and Prevention Contract #75D30120P07909; Centers for Disease Control and Prevention Contract #75D30121P10548; National Science Foundation 20-052 – COVID-19 Rapid Response Research (RAPID)

Disclaimers: The findings and conclusions in this publication are those of the authors and do not necessarily represent the views of the North Carolina Department of Health and Human Services, Division of Public Health, the Centers for Disease Control and Prevention, or the National Science Foundation.

# Table of Contents

**Introduction**

This agent-based model (ABM) has taken many forms during its lifetime. It was originally designed to simulate interventions to help reduce transmission of infectious diseases, such as *Clostridioides* difficile infection[1,2] and carbapenem-resistant Enterobacteriaceae. In 2020, it was adapted to model hospitalizations during the COVID-19 pandemic.[3] Each application of the underlying agent movement model (further referred to as the *base model)* led to adjustments, updates, and expert opinions affecting how agents moved to and from healthcare facilities. Keeping these updates scattered across multiple code repositories and ODD documents became difficult as additional models were created. To make model improvements available to all future applications of the base model, we have collected these changes and created a robust movement model that is independent of all submodels. The base model is no longer attached to specific disease submodels. However, it can be controlled by individual submodels as needed. This ODD is a living document. As updates are made in the code repository, the ODD is subject to change.

**1. Purpose and Patterns**

The purpose of this ABM is to create a realistic simulation of agent movement to and between all modeled healthcare facilities in North Carolina to be used as a base model for disease-specific submodels. To validate how realistic this base model is, the model's patterns include matching known length-of-stay (LOS) values, individual facility capacities, transfers between different facility types, and agent demographics for agents in each type of facility. By accurately matching historical information about facilities, we can simulate the impact of disease-specific interventions, changes in hospitalizations, and other scenarios related to the healthcare system in North Carolina.

*ABM Patterns*

To validate agent movement, we evaluate the ABM by its ability to reproduce several patterns. To compare model output to expected values, the ABM is run one time, using 1-year time horizon and over 10 million agents (the population of North Carolina). Since the number of modeled steps (365 days), the agent population, and the pattern targets are all large values, it is not necessary to run multiple simulations when reviewing these patterns. All model results are reproducible by setting the model's random seed.

**Pattern 1: Length of stay.** We compare length of stay (LOS) values assigned during a model run to the input LOS distribution for each facility. Table 1 is the output for the four largest hospitals (by admissions) in the model. We show the modeled and expected value for admissions, average LOS, and the LOS standard deviation. Values for all facilities are available in the code repository. For this pattern, the modeled and expected values should be almost identical.

*Table 1. Example model output for pattern 1*

| Facility ID | 365-day Admissions | Average LOS | Expected average LOS | Sd. LOS | Expected Sd. LOS |
|---|---|---|---|---|---|
| 23 | 49,304 | 5.72 | 5.77 | 2.48 | 2.5 |
| 31 | 37,448 | 4.91 | 4.96 | 2.13 | 2.13 |
| 66 | 36,942 | 6.02 | 6.07 | 2.66 | 2.64 |
| 89 | 36,151 | 7.21 | 7.29 | 3.21 | 3.21 |

**Pattern 2: Average capacity.** We compare the average capacity value for each facility over the duration of a model run to the average capacity specified as input for each facility (Table 2). We also check that average capacity is consistent over the model run and does not steadily increase or decrease over time for each facility. For this pattern, we expect all large facilities to be within 5% of the expected value.

*Table 2. Example model output for pattern 2*

| Facility ID | Average Capacity | Min | Max | Sd. | Expected Capacity |
|---|---|---|---|---|---|
| 23 | 885.14 | 809 | 960 | 25 | 894 |
| 36 | 740.18 | 677 | 713 | 25 | 738 |
| 89 | 738.26 | 678 | 799 | 23 | 723 |
| 62 | 639.72 | 578 | 692 | 21 | 629 |

**Pattern 3: Agent movement between healthcare locations.** We aggregate agent movement between different facility types and compare these values to aggregate North Carolina discharge data.[4] To do this, the number of all movements from one location type to another is recorded and compared to the expected number of movements for those two location types (Table 3). Movement between certain location types (e.g., community to community) is not possible. For this pattern, we expected all larger targets to be within 5%.

*Table 3. Example model output for pattern 3*

| Location Type | New Location Type | Target | Modeled Value |
|---|---|---|---|
| Community | Community | 0 | 0 |
| Community | Hospital | 825,150 | 815,414 |
| Community | Nursing home | 21,851 | 21,592 |

## 2. Entities, State Variables, and Scales

The base model has two types of entities—agents and locations. Agents represent North Carolina residents informed by a synthetic population of North Carolina. There is one agent per resident. Locations include over 500 nodes that are part of a geospatially explicit network. These nodes represent healthcare facilities, where agents will seek healthcare during a model run, and the community (Table 4, Figure 1). Agents moving between these nodes represent North Carolinians moving among healthcare facilities and the community. There are four location types: short-term acute care hospitals (STACHs), long-term acute care hospitals (LTACHs), nursing homes (NHs), and the community.

Table 4. Location type and count of facilities in the ABM

| Location Type | Description | Node Count |
|---|---|---|
| STACH | Short-term acute care hospitals[5,6] | 104 |
| LTACH | Long-term acute care hospitals[6] | 10 |
| NH | Nursing homes[7] | 426 |
| Community | Everything outside of the healthcare facility types | 1 |

Agent movement among households is not modeled, nor is agent interaction within a household. Agents located in the community node can be conceptualized to be anywhere in the community other than a modeled STACH, LTACH, or NH. Healthcare facility locations (i.e., all noncommunity nodes) have static variables, including name, physical location (i.e., county and geocode), bed count, and a unique identifier (Facility ID) that is used throughout the ABM (Table 2). Agents who move to one of these locations are given a specific bed within the healthcare facility. At that time, an agent is added to the "Agents" attribute of the location node. STACH nodes have beds designated as intensive care unit (ICU) or non-ICU beds. Not all North Carolina counties have a modeled healthcare facility. Note that most variables are not relevant to the community node.

Table 5. Location node specific variables

| Variable | Description | Dynamic | Type | Range |
|---|---|---|---|---|
| Bed Count | non-ICU and ICU | No | Integer | 1-1000+ |
| Name | Healthcare facility name | No | String | N/A |
| Category | Healthcare facility category | No | String | N/A |
| County | North Carolina county of the healthcare facility | No | String | 1, 3, 5, … 201 |
| Agents | Current agent IDs | Yes | List | N/A |
| Facility ID | Unique facility identifier | No | Integer | 0-540 |

Each time step (i.e., 1 day), an agent's location state can be updated. The location state corresponds to the current location for that agent. An agent's life state is a binary variable (i.e., living, or dead). Agents also have several demographic attributes. Throughout the model run, agents who change locations are assigned additional variables. An agent's LOS, leave healthcare facility day (model date on which the agent's LOS ends if they are at a facility node), previous location, other information may be added, updated, or removed throughout the model run. During model initialization, agents are also assigned to presence or absence of comorbidities, labeled as "concurrent conditions" in the model (see **Section 5: Initialization**).

Table 6. Agent state and demographic variables of the ABM

| Variable | Description | Dynamic | Type | Range |
|---|---|---|---|---|
| Unique ID | ID for the agent | No | Integer | 0-10,600,822 |
| Age Group | Age of the agent (<50 (0), 50-64 (1), 65+ (2)) | No | Integer | 0, 1, 2 |
| County | Home county | No | Integer | 1-201 |
| Concurrent conditions (i.e., comorbidities) | Binary variable for presence of comorbidities | No | Integer | 0, 1 |
| Location | Current location | Yes | Integer | 0-540 |
| Life | Life status | Yes | Integer | 0, 1 |

*Temporal and Spatial Resolution and Scales*

The ABM is implemented with a 1-day time step, and there is no sense of daily time in the model. A 1-day time step was selected based on the assumption that an agent is primarily at one location each day. Submodels can have more granular time steps as needed. The number of daily time steps in the model is set by an input parameter. Each facility in the model is geospatially explicit based on the GPS coordinates of the facility's address. Agents in the model are also geospatially explicit because they are from a specific North Carolina County. Each county is described by the coordinates at the county's centroid. This spatial resolution was chosen based on data availability and being able to move agents to locations based on the distance between an agent's home county and each facility.

## 3. Process Overview and Scheduling

For each time step, agents can go through several different updates, we will call these updates actions. Any action that was randomly selected to occur for the current model time, is placed in a list of actions for the model to execute. These actions contain the agent's unique ID and the type of action to execute. Examples of actions include an agent seeking hospitalization from the community, being transferred from one facility to another, or dying at a facility. Randomizing the updates before they are executed is important. Resources in the model (e.g., beds) may be limited. Randomizing the actions allows resources to be used in a more realistic manner. Randomization also helps maintain a realistic simulation of an agent's day. In the base model, only location movement and life actions occur. As additional submodels are used, there will be additional agent actions that can take place. All actions should be described in the submodel section of accompanying ODDs.

## 4. Design Concepts

There are several design concepts available in an ODD protocol. We have briefly described a few of them below and will continue to add to this section as needed.

*Basic Principles*

To be added.

*Emergence*

Model results because of emergence will primarily be dependent on the disease-specific submodel that uses the base model. The patterns related to agent movement to and from facilities is driven by strict transition probabilities.

*Adaptation*

Agents can be turned away from facilities if a facility is at capacity. They may choose a new facility to go to, but these decisions are rule based. Adaptation may be more relevant when a disease-specific submodel is implemented.

*Objectives*

To be added.

*Prediction*

The ABM does not currently have explicit or implicit prediction components.

*Sensing*

Agents have no concept of knowledge; therefore, sensing is not relevant.

*Interaction*

Agents in the ABM only have mediated interactions. When an agent occupies a bed in a healthcare facility, other agents do not have access to that resource. This might cause agents to be turned away from an STACH. Disease-specific submodels might have additional agent interactions.

*Stochasticity*

To simulate random events that happen to each agent (i.e., going to a hospital), the ABM compares the probability of events to values generated by a random number generator. If the probability of an event is greater than the random number, the event is placed in the queue of actions to execute. The order that actions are executed is also randomized. By setting a random seed, results can be reproduced.

*Collectives*

To be added.

*Observation*

To be added.

**5. Initialization**

*Agents*

Agents for the ABM are created using demographic variables provided by RTI SynthPop™.[8] This synthetic population consists of one anonymized synthetic person per row, containing the synthetic person's home county; Federal Information Processing Standard code (corresponding to one of North Carolina's 100 counties); sex (female, male); and age in years. We completed two preprocessing steps on this file to prepare it for model input:

1. This baseline population is slightly smaller (~10m) than the estimated population of North Carolina for 2020 of 10,600,823.[9] Rows in the synthetic population are randomly selected and duplicated until we reach the population estimate of North Carolina for 2020. Only agent demographic information such as age, sex, and general location are used; duplicating agents is acceptable here because no household-specific attributes (e.g., household size) are used.
2. We binned the age of the agents into age groups (<50, 50-64, ≥ 65) to match the age groups used in previous models.

A base model parameter sets the number of agents to include in the model. This is often much smaller than the full population of North Carolina. Using a smaller population reduces the amount of time it takes a model to complete, but also reduces the accuracy and consistency of the results. Based on this parameter, rows from the synthetic population are randomly selected and an agent is initiated with the values of that row. Each agent has the same probability of being selected. Unless selected to start in a facility, all agents initially start in the community. Agents are assigned two additional variables outside of the variables in the synthetic population:

1. Agents are given a unique id. This value is used to track their movements and record any events that take place.
2. Agents are randomly assigned concurrent conditions (i.e., comorbidities) (Table 4). This is a binary variable that is based on their age group.[10]

Table 7. Probability of concurrent condition assignment by age group

| Age Group | Probability |
|---|---|
| 0 (<50) | 0 |
| 1 (50-64) | 23.74% |
| 2 (65+) | 54.97% |

*Location Nodes*

Location entities are created using input data that provide healthcare facility names, locations, bed counts, and other attributes (see **Section 7. Input Data**). Each facility is assigned a specific number of beds based on the input data and the number of agents in the model. The number of beds for facility $i$, $B_i^*$, is equal to the number of beds multiplied by the ratio of agents in the model, $n$, to the population of North Carolina, $p$. Each facility must have at least one bed.

$$B_i^* = \max(1, B_i * \frac{n}{p})$$

To fill these beds, starting capacities are determined by input parameters for STACHs and LTACHs, or facility-specific capacity estimates for nursing homes (Table 8).[7] These values can be changed to better suite scenario-specific setups, and facility-specific values can be used for STACHs and LTACHs, if available.

Table 8. Parameters for starting capacities of STACH and LTACH facilities

| Parameter | Description | Value | Source |
|---|---|---|---|
| LTACH fill | Proportion of LT beds filled at initiation | .9 | Opinion |
| Non-ICU fill | Proportion of non-ICU hospital beds filled at initiation | .65 | Opinion |
| ICU fill | Proportion of ICU hospital beds filled at initiation | .50 | Opinion |

Each STACH is initiated with ICU and non-ICU agents to match starting capacity percentages specified by model parameters. Alternatively, in cases where individual hospital capacity data are available from another source, the model can initialize non-ICU and ICU capacities to the provided levels. Parameters control whether a global fill value is used or if real data are used. When using input parameters, STACHs will not all start at the same capacity level. Instead, they will start at capacities that reflect both their reported discharge data and the input parameter. On average, hospitals will start at capacities equivalent to the fill input parameters for non-ICU and ICU beds. The following is used to estimate hospital-specific starting capacities. Let $H_{NB}$ be the number of non-ICU beds, $H_{TB}$ be the total number of beds, $H_{Tp}$ be the total number of discharged patients, and $H_{LOS}$ be the mean LOS for hospital $H$. $H_c$ is the average capacity for non-ICU beds of a facility based solely on discharge data.[4,5]

$$H_c = \frac{H_{NB}}{H_{TB}} * H_{TP} * \frac{H_{LOS}}{365}$$

Let $R$ be the ratio of the input parameter, *non-ICU fill,* and the estimated average capacity of all hospitals. Then $H_c^*$, the actual starting capacity of hospital $H$ for non-ICU beds given the discharge data and the input parameter, is $R * H_c$. A similar process is completed for ICU beds. We do not model agents that are not North Carolina residents. However, several hospitals have large amounts of individuals from states other than North Carolina. Some hospital beds will be assigned *placeholder agents*. These placeholder agents represent agents at each facility that are not North Carolina residents. Placeholder agents do not move and do not have attributes and the total number at each facility remains constant for a model run. The number of placeholder agents is determined by multiplying the capacity of a facility by the proportion of discharges that were not North Carolina residents.

Agents are selected to fill hospital beds based on their home county and the discharge data for each hospital.[5] The probability that a bed will be filled by an agent from a specific county is equal to the number of discharges from that hospital and county divided by the total number of discharges from that hospital. After counties have been selected, agents within those counties are chosen based on their age (Table 9).[4]

*Table 9. Probability of hospitalized agent being from a specific age group*

| Age Group | Probability |
|---|---|
| 0 (<50) | 40.99% |
| 1 (50-64) | 20.12% |
| 2 (65+) | 38.90% |

All agents who were assigned a starting location other than the community are assigned a *remaining LOS* value. This value is drawn from a distribution created by aging facility-specific LOS distributions. This process is completed before ABM execution and is solely used to create a *remaining LOS* distribution. This aging process consists of continuously pulling values from the normal LOS distribution as days are simulated. For each day that passes, 1 day is removed from the drawn values. After enough time has passed for the distribution to reach a steady state, whatever values remain that are greater than 0 are used as the remaining LOS distribution. These distributions are only used during model initialization.

*LTACHs*

For LTACHs, all facilities will be assigned agents based on the fill parameter and the number of LTACH beds at that facility. Counties are selected with replacement using, $P_{ij}$, the probability that a person in facility $i$ is from county $j$. Probabilities are based on the distance between the facility and the center of the county.

$$P_{ij} = \frac{\frac{1}{D_{IJ}}}{\sum D_{ij}}$$

$D_{ij}$ is the distance between facility $i$ and county $j$. Therefore, counties closer to a facility are more likely to be selected. Counties are selected equal to the desired number of beds. Agents from selected counties are then randomly selected. For an LTACH, agents must be 50+. Selecting between age groups 1 and 2 is not weighted by age.

*Nursing Homes*

Nursing home facilities are initiated with agents based on facility-specific capacity values.[7] Agents are randomly selected based on a probability distribution formed by taking the relative distance between the facility and each county, as was done in the LTACH initiation. For Nursing homes, agents must be ≥ 65 years of age to be selected.

## 6. Input Data

Below we describe the different data files used in the ABM. After the title of the file, we have provided the location in the code repository where the data file is located.

*County-Distances (data/geography/…)*

There are three geography files that must be included, one for each facility type. These files consist of the distance from each county center to the geocode for each facility in miles, and they are used to help select new facilities for agents. The code used to automatically create these files is available in the code repository.

*Locations (data/locations/…)*

To create appropriate facilities within the model, a file for each facility type is required. The hospital file (STACHs) consists of hospital name, location, and the count of ICU and non-ICU beds. It was extracted from an online list of North Carolina hospitals.[11] However, only hospitals that have information in the discharge data will be used. The nursing home file contains facility name, bed count, county, and geocode information. The nursing home facilities were taken from

Centers for Medicare & Medicaid Services (CMS) data[7] and geocodes were programmatically added. The file for LTACHs was also derived from online data and contains the name, bed count, and geocode for each facility.[11]

*Discharge Data (data/sheps_data/2018/…)*

PDFs of the public North Carolina hospital data files are available in the repository. Data were automatically extracted from these PDFs and converted to the CSV files that the model reads as input. Code and instructions for converting PDFs to CSVs is in the code repository. There are three files:

1. *Patient county of residence by hospital* provides the list of counties that agents came from for each hospital[5]

2. *Short term acute care hospital patient characteristics* provides details on patient age group, home state of the patient, and patient disposition[4]

3. *Short term acute care hospital discharge data* provides an LOS estimate for each facility[12]

*Synthetic Population (data/synthetic_population/synthetic_population.parquet)*

The synthetic population file provides agents for the model. It was provided by RTI International.[8] Two updates to this file are described in **Section 5 Initialization**.

**Data Calculated at Initiation**

Before location and agent entities are created, model transition probabilities and additional required input are calculated based on input parameters and the input data files. This preparation script is available in the repository. We provide brief details on each component below. There are several additional parameters used to create these data.

*Table 10. Additional location movement parameters*

| Parameters | Description | Value | Source |
| --- | --- | --- | --- |
| Nursing home Death | Proportion of nursing home agents who die each year | .15 | Opinion |
| LTACH - Hospital | Proportion of LTACH agents moving to an STACH upon discharge | .071 | [13] |
| LTACH - NH | Proportion of LTACH agents moving to a nursing home upon discharge | 0.449 | [13] |
| LTACH Death | Proportion of LTACH agents who die (simulated at discharge) | 0.01 | [13] |
| LTACH 65+ | LTACH agents who are 65 plus | 0.75 | Opinion |
| NH - STACH - NH | Proportion of nursing home agents who return to a nursing home after a hospitalization | 0.80 | [13] |
| NH - Community | Proportion of nursing home residents who move to the community at discharge | 0.67 | Opinion |

*Component: hospital-df*

This object merges all known information about each hospital across the different input files and provides the starting capacities for the model.

*Component: community-transitions*

For each county and age group combination, daily probabilities are calculated to represent the likelihood that agents in the community with those characteristics will seek treatment in a hospital or a nursing home. LTACH probabilities are not included because community agents cannot go directly to an LTACH. These probabilities are based on how many agents with those characteristics were admitted to each facility type from the community each year.

*Component: facility-transitions*

For each individual hospital, as well as a collective category for nursing homes (representing all nursing homes) and a collective category for LTACHs (representing all LTACHs), we calculate the probability of an agents discharge location when their LOS is over. The probabilities are split between community, hospital, LTACH, and nursing home and each row must sum to 1; they are broken down by age. The probabilities are based on how many people left that facility or facility type and were discharged to each facility type for the data available. The probabilities are adjusted so that agents less than 65 years old cannot enter a nursing home, and agents less than 50 years old cannot enter an LTACH.

*Component: death-dictionary*

For each facility type, we calculate the probability of death based on input parameters. This probability is equal to the number of deaths each year at each facility type divided by the number of discharges. Death is simulated when an agent's LOS ends.

*Component: four-by-four*

During initiation we also calculate the targets for the model run. The four-by-four is a matrix of the four facility types with each row representing the number of agents that should move from one facility type to the next.

*Component: hospital-age-distribution*

Based on input data, we calculate the distribution for the age of agents that go to hospitals.

## 7. Submodels

The ABM has two submodels, one for life and one for location. Additional submodels can be appended as needed.

*Life*

Death is an important component of the ABM. Agents who die at a facility are not able to transfer to another facility, which allows the model to maintain proper capacities. Death only occurs when an agent's LOS ends, and death probabilities are set using input parameters and discharge data.[4] The probability of death for hospitals is based on discharge data and is calculated as the number of deaths at hospitals divided by the number of total discharges. LTACH and nursing home death probability is based on model input parameters. The life submodel is meant to be controlled by additional submodels as well. Agents can die from submodels that use the base model.

*Location*

The location submodel controls agent movement. Agents move locations if one of these four situations occurs: (1) agents in the community are randomly selected to move based on their daily probability of movement; (2) agents who previously left a healthcare facility moved to the community and were given a readmission date to return to that healthcare facility on their readmission day (this is currently not active in the model); (3) agents in healthcare facility nodes move when their LOS ends; or (4) a submodel moves agents based on disease spread or other factors.

*Location: community movement*

Each agent in the community has a daily probability of leaving the community and going to either an STACH or a nursing home. In the community movement step, the probabilities of movement for each agent are compared to random numbers to see if that agent will leave the community. If selected, an action for that agent is placed in the action queue depending on which type of movement was selected (community to STACH or community to nursing home). When the action is executed, the agent will be assigned to a specific location and given an LOS. This "first-choice" location is based on an agent's home county and how often facilities admit patients from that county.

*Location: facility movement*

Agents in healthcare facilities only leave that facility when their LOS ends. We use the location transition probabilities to determine the healthcare facility type of the agent's next destination or to determine if the agent dies. A random probability is generated, and this probability is compared to their transition probabilities. Most agents will move to the community upon discharge, but some are selected for a facility transfer. Once a healthcare facility type is determined, if a non-community node is selected, we compare a second random number to the facility transitions to determine the exact facility ID. For agents who previously transitioned from a nursing home to an STACH, we assume that most of these agents will return to a nursing home. This probability is set by an input parameter. We included this hard-coded component in the face of a general lack of available data on nursing home agent movement; this can be updated in the future with additional data.

Anytime an agent moves to a new STACH, they have a probability of requiring an ICU bed. This probability is based on patient-level data previously used to calculated LOS distributions.[1] A logistic regression model was built using an agents age, concurrent conditions, and selected LOS, as well as the hospital's bed count to determine the probability of requiring an ICU bed. This probability is also controlled by an input parameter that can increase or decrease ICU stays. This input parameter was calibrated to create a steady state of ICU patients over the course of a 1-year model run.[1]

When an agent arrives at an LTACH, the agent is assigned an LOS based on a gamma distribution. For agents that arrive at an STACH or a nursing home, we use facility-level distributions based on input data.

*Location: facilities at capacity*

It is possible that a facility will be at capacity when an agent tries to seek treatment. The following set of rules are applied when an agent goes to a facility that is at capacity:

1. If the facility is a hospital, the agent will try to find any open bed, regardless of whether it is an ICU or non-ICU bed, and which type the agent initially sought.

2. The agent will try to find a bed at all facilities located in their home county.
3. The agent will try any additional North Carolina facilities located within a 200-mile radius of the centroid of the agent's home county; this value is controlled by an input parameter.

If the agent is turned away from their first-choice facility, that agent is added to a list of agents turned away. This list contains a record for each agent turned away, and it includes the date, location, and county. If the agent is turned away from all possible facilities, that agent is added to a list of agents who were completely turned away during the model run. Agents who are randomly selected to transfer from another healthcare facility will only try their first-choice facility. If this STACH is at capacity, the agent returns to the community. We based this assumption on the premise that a healthcare facility would not transfer an agent to a facility that did not have a bed available for them (Table 11).

*Table 11. Distance Parameters*

| Parameters | Description | Value | Source |
|---|---|---|---|
| Nursing home Closest N | Number of nursing homes considered when an agent is turned away because of capacity issues | 30 | Opinion |
| Nursing home Attempts | Number of nursing homes an agent will try to go to before giving up | 30 | Opinion |
| LTACH Closest N | Number of LTACHs considered when an agent is turned away because of capacity issues | 10 | Opinion |
| LTACH Attempts | Number of LTACHs an agent will try to go to before giving up | 3 | Opinion |
| Max. Distance | The maximum number of miles a facility can be located from an agent's home county | 200 | Opinion |

Acknowledgements: We are grateful for the support and input from Susan Eversole, Stacy Endres-Dighe, Georgiy Bobashev, and Alex Giarrocco of RTI International and from our UNC

Health collaborators, and our public health collaborators. This activity was based on a model originally developed through support from CDC's Modeling Infectious Disease in Healthcare (MInD-Healthcare) Network.